\documentclass[runningheads]{llncs}
\usepackage[T1]{fontenc}
\usepackage{graphicx}
\usepackage{subcaption}

\usepackage{csquotes}
\usepackage{paralist}
\let\oldFootnote\footnote
\newcommand\nextToken\relax

\renewcommand\footnote[1]{%
    \oldFootnote{#1}\futurelet\nextToken\isFootnote}

\newcommand\isFootnote{%
    \ifx\footnote\nextToken\textsuperscript{,}\fi}

\usepackage[breaklinks]{hyperref}
\expandafter\def\expandafter\UrlBreaks\expandafter{\UrlBreaks%
  \do\a\do\b\do\c\do\d\do\e\do\f\do\g\do\h\do\i\do\j%
  \do\k\do\l\do\m\do\n\do\o\do\p\do\q\do\r\do\s\do\t%
  \do\u\do\v\do\w\do\x\do\y\do\z\do\A\do\B\do\C\do\D%
  \do\E\do\F\do\G\do\H\do\I\do\J\do\K\do\L\do\M\do\N%
  \do\O\do\P\do\Q\do\R\do\S\do\T\do\U\do\V\do\W\do\X%
  \do\Y\do\Z}
\usepackage[nolist]{acronym}
\usepackage{listings}
\usepackage{amssymb} %

\usepackage{booktabs}
\usepackage{tabularx}

\usepackage{silence}
\usepackage{caption}
\usepackage{subcaption}

\newcommand{\ph}{Provenance Holder}

\usepackage{makecell}

\usepackage{xcolor}

\begin{document}
\title{AdProv: A Method for Provenance of Process Adaptations}
\author{Ludwig Stage\inst{1}\orcidID{0009-0005-4431-2203} \and
Mirela Riveni\inst{1}\orcidID{0000-0002-4991-3455} \and \\
Raimundas Matulevi\v{c}ius\inst{2}\orcidID{0000-0002-1829-4794}\and \\
Dimka Karastoyanova\inst{1}\orcidID{0000-0002-8827-2590}}

\authorrunning{L. Stage et al.}
\institute{University of Groningen, The Netherlands\\ 
\email{\{l.stage, m.riveni, d.karastoyanova\}@rug.nl}\\
University of Tartu, Estonia\\
\email{raimundas.matulevicius@ut.ee}}
\maketitle              %
\begin{acronym}
	\acro{ACID}{Atomicity, Consistency, Isolation, Durability}
	\acro{BPEL}{Business Process Execution Language}
	\acro{BPM}{Business Process Management}
	\acro{BPMN}{Business Process Model and Notation}
	\acro{BPMS}{Business Process Management System}
	\acro{EPR}{Endpoint Reference}
	\acro{ESB}{Enterprise Service Bus}
	\acro{FAIR}{Findable Accessible Interoperable Reusable}
    \acro{KPI}{key performance indicator}
        \acro{MitM}{Machine-in-the-Middle}
        \acro{PAIS}{Process-Aware Information System}
    \acro{PROV-O}{PROV Ontology}
	\acro{PoC}{Proof of Concept}
	\acro{SLA}{Service-level agreement}
	\acro{SOA}{Service-oriented Architecture}
	\acro{SOC}{Service-oriented Computing}
	\acro{RARE}{Robust Accountable Reproducible Explained}
   \acro{UDDI}{Universal Description, Discovery and Integration}
   \acro{UUID}{Universally Unique Identifier}
   \acro{WSDL}{Web Services Description Language}
   \acro{WfMS}{Workflow Management System}
   \acro{sWfMS}{Scientific Workflow Management System}
\end{acronym} \begin{abstract}
Provenance in scientific workflows is essential for understanding and reproducing processes, while in business processes, it can ensure compliance and correctness and facilitates process mining. However, the provenance of process adaptations, especially modifications during execution, remains insufficiently addressed.
A review of the literature reveals a lack of systematic approaches for capturing provenance information about adaptive workflows/processes. To fill this gap, we propose the AdProv method for collecting, storing, retrieving, and visualizing provenance of runtime workflow adaptations.
In addition to the definition of the AdProv method in terms of steps and concepts like change events, we also present an architecture for a \ph{} service that is essential for implementing the method. To ensure semantic consistency and interoperability we define a mapping to the ontology \ac{PROV-O}. Additionally, we extend the XES standard with elements for adaptation logging. %
Our main contributions are the AdProv method and a comprehensive framework and its tool support for managing adaptive workflow provenance, facilitating advanced provenance tracking and analysis for different application domains.

\end{abstract}

\section{Introduction}
\ac{BPM} is a mature field that offers concepts and technologies covering the complete life cycle of processes/workflows\footnote{The terms processes and workflows are used here as synonyms and denote processes managed by a computing environment, unless explicitly distinguished.} and enables Process Systems that can be used to specify and model processes, execute their instance, monitor them and collect historical data in the form of e.g. process event logs. The analysis of process event logs helps improve the processes with respect to \acp{KPI} of interest, identify potential compliance issue and discover deviations from predefined process models. 

With the advent of ML, Data science and AI, process mining in general and predictive process monitoring, in particular, have been instrumental in providing means to detect reasons/triggers for potential changes in running processes, the so called adaptations. Adaptations may also be dictated by changes in legislation, regulations, company related policies and other circumstances. 
In literature, we distinguish between evolutionary changes that affect the process models and hence all their instances and ad-hoc or runtime changes that happen during process instance execution and affect only some instances \cite{WeberReichertonAdaptationBook}. 
Process systems research is mature enough to enable evolutionary and ad-hoc changes already. 

What is still missing however, as our literature review shows (see Section \ref{sec:literature}), are methods and process systems that systematically capture the changes made on processes as first class citizens (actions of interest), identify and record them, and enable their management and meaningful use and visualization. 
In \ac{BPM} such methods and the enabling software services would allow for checking and ensuring correctness of changes and compliance to predefined models and regulations, would allow to protect the security and privacy properties of processes from potential violations in case of adaptations and furthermore be an enabler of new methods for learning from adaptations for faster and more efficient process change.

The field of scientific workflows has delivered a number of systems and tools that automate the data processing steps carried out in scientific research and discovery. These systems typically focus on optimal scheduling of data processing steps forming so called experiments, typically but not always, for specific research domain.
Scientific workflow management systems are built to support the trial-and-error nature of scientific discoveries which means that in many cases scientific workflow models are consonantly being modeled while being executed, which is in stark contrast to the conventional workflow technology from \ac{BPM} (where process  modeling and execution are kept separate). Furthermore, scientific workflow management systems serve smaller number of users than conventional process systems from \ac{BPM} and do not deal with fault handling part of the workflow  models. However, they put special emphasis on provenance of the workflows and the data used to enable FAIR\footnote{Findable Accessible Interoperable Reusable } and RARE\footnote{Robust Accountable Reproducible Explained} research and guarantee reproducibility of research results and ensure their quality. Therefore, as our literature review shows, provenance information is typically collected and standards for provenance data models have been established. Similarly to the \ac{BPM} field, provenance of adaptations has not been considered as first class citizen and a method for its collection, storage and management has not been engineered.

In response to the lack of support for managing provenance of process adaptations, our \textit{contributions} in this work are:
\begin{compactitem}
    \item An overview of the state of the art of provenance of adaptation based on two systematic literature reviews (SLRs)
    \item The AdProv method for collecting, storing, retrieval and visualization of provenance of workflow adaptations
    \item A data model for capturing changes in so-called change events and a corresponding extension to XES
    \item A mapping of process execution and change events to the provenance standard \ac{PROV-O} to support meaningful provenance information visualization
    \item Refined architecture of the \ph{} service accommodating the concepts of the AdProv method
     \end{compactitem}

We present our contributions along the following paper structure. The state of the art is presented in Section \ref{sec:literature}. Section \ref{sec:method} introduced the AdProv method while Section \ref{sec:realization} is devoted to the definition of change events, the Adaptation XES extension, the \ph{} architecture and the mapping of provenance information to the PROV-O model for the purposes of visualization. We discuss the method feasibility in Section \ref{sec:feasibility} and conclude the paper with  Section \ref{sec:conclusions}.

\section{State of the Art}
\label{sec:literature}

In this section, we provide a brief overview of the related work findings based on two systematic literature reviews (SLRs) following a standard methodology ~\cite{Kitchenham2007} for establishing the current state of the art. We considered the available literature on two main topics: a) the available solutions for provenance of runtime changes in workflows and b) the existing approaches for logging runtime workflow adaptations and process mining approaches on such process event logs.

\textbf{Workflow Provenance and Provenance of Workflow Adaptation:}

In \cite{stage2025provenanceadaptationscientificbusiness} we report the details of the SLR and our findings on the state-of-the-art of \textit{provenance of workflows adaptations} in both business and scientific workflows. 

As a result of the SLR methodology we selected 22 publications from the initial 1165 discovered in four literature libraries. It is worth noting that only 5 papers are from the BPM community, and only 8 papers mention provenance of adaptation or synonyms. %
We observe that the \textit{information} that is being \textit{captured} focuses mainly on workflow inputs, outputs and their models with different granularity and purpose. The only mentions of evolution provenance is in \cite{id:20_herschel_2017} which states that Kepler \cite{kepler} and VisTrails \cite{vistrailswiki,id:12_2006} provide support for provenance of process evolution for scientific workflow systems in which there is no decoupling between the process model and its instances. %

Typical \textit{use cases of and motivation} for provenance are fraud prevention to ensure trust, reproducibility, analysis and validation, quality control, error handling, development. %

\autoref{imp-tools} summarizes the findings regarding the available workflow provenance tools.
All these tools originate from the scientific workflow community and are open source. Only VisTrails supports most of the necessary capabilities and can collect/capture, manage and visualize provenance information. %
According to \cite{id:20_herschel_2017}, evolution provenance is supported by VisTrails and Kepler, but none of the tools are reported to support ad-hoc provenance. 

\renewcommand{\arraystretch}{1.0}
\begin{table}
\centering
\caption{Workflow provenance tools.
C: Capture, M: Manage, V: Visualize, E: Edit}\label{imp-tools}
\makebox[1 \textwidth][c]{%
\resizebox{1.0 \textwidth}{!}{%
\setlength{\tabcolsep}{0.11cm}
\begin{tabular}{lccccc}
\toprule
\textbf{Name} & \textbf{Purpose} & \textbf{Standard} & \textbf{Type} &  \textbf{Last Change} \\
\midrule
Kepler \cite{Arxiv2023,id:16_2022,id:15_2008,id:9_2021,id:4_2008,id:3_2017,id:20_herschel_2017}%
& C & PROV, OPM & \ac{WfMS} &  2021 \\

Prov Viewer \cite{Arxiv2023,id:22_2021}%
& V & PROV & Standalone &  2020   \\

ProvViz \cite{Arxiv2023,id:22_2021}%
& V, E & PROV, RDF & Standalone & 2021  \\

ReproZip \cite{id:10_2017}%
& C, M & None & Standalone & 2024  \\

Taverna \cite{id:9_2021,id:4_2008,id:3_2017,id:18_2019,id:20_herschel_2017}%
& C, M & PROV, OPM & \ac{WfMS} &  2020   \\

VisTrails \cite{id:10_2017,id:19_avocado_2016,id:12_2006,id:9_2021,id:4_2008,id:20_herschel_2017,id:21_2016}%
&  C, M, V & OPM & \ac{WfMS} & 2017  \\

\bottomrule
\end{tabular}
}
}
\end{table}

In the same SLR we also present an overview of conceptual methods and when relevant corresponding tool prototypes. 

Based on this overview we conclude that the main objective of most of the tools is to only capture and record the provenance information of workflows and only some are supporting management and visualization of such information. Only two solutions indicate some form of support for provenance of adaptation: PRISM and (our own) Provenance Holder. %
The interested reader is referred to \cite{stage2025provenanceadaptationscientificbusiness} for more details.
 
PRISM's  main goal is to enable coordinated adaptations, which in addition can be recorded as part of the provenance information it gathers. Despite using blockchain to ensures complete and transparent provenance records, the provenance of workflow adaptations in PRISM is merely a by-product and cannot be applied with other WfMS or PAIS. 

The Provenance Holder \cite{Arxiv2023} is a service designed with the purpose to support all forms and granularities of workflow provenance (\cite{stage2025provenanceadaptationscientificbusiness}) and can be used by different WfMS and PAIS. 
The Provenance Holder is designed to collect provenance on a very detailed level during the execution and adaptation of workflows following the AdProv method presented in this work.

\textbf{Adaptation Logging and Mining Workflow Logs:}
In \cite{2025changeloggingminingchange} we report our finding from an SLR on \textit{recording and logging of runtime workflow changes as well as on the mining of such logs} in PAIS. 
Our goal was  to identify the existing approaches for logging and mining of process change in business processes and workflows in PAIS and included the areas of service compositions.%

Based on the selected publications we observe that \textit{changes} in business processes \textit{are recorded} into change logs either by recording each change operation carried out by a PAIS \cite{waas} or identified by comparing original and revised process models \cite{waas} and recorded in a next step. 
Recording workflow changes at runtime carried out by a PAIS is reported in only two publications (\cite{waas} and \cite{purging}).
It is noteworthy that in most reported cases change information does not get recorded at all and has to be discovered from event logs (e.g. \cite{anewframework}). 

The selected publications report on two \textit{main categories of methods for change mining}: 
i) Adapted Process Mining Methods -- these are methods that apply and/or extend existing process mining methods for change mining purposes. Application of $\alpha$-Algorithm, MultiPhase Miner, and Heuristics Miner algorithms to MXML change logs are available in this category. %
ii) Novel Methods for Change Mining that are not extensions of existing process mining methods. The methods in this class are: mappings of change logs to labeled state transition systems \cite{usingprocessmining}, algorithms based on machine learning \cite{handlingsudden}, and methods based on comparing process structure trees \cite{waas}.
We also identified four categories of \textit{results produced by the  mining approaches}:
change models \cite{usingprocessmining}, change trees \cite{applyingchange,miningquerying}, change recommendations \cite{anewframework} and change rules ~\cite{waas}.

The findings show that the focus of the BPM community has been on change detection approaches for the purpose of process improvement but the literature does not show evidence that provenance of changes has been in the primary focus of BPM research in systematic way.

Overall, the related literature shows that provenance of workflow adaptations of any granularity and type has not been in the focus of the BPM and scientific workflow communities. Moreover there is a lack of methods that would enable the development of systems for gathering and using provenance information about workflow adaptations. In the next section we will introduce one such method.

\section{The AdProv Method}
\label{sec:method}
We use the knowledge from method engineering as it serves our objective to systematically develop and introduce a new “method” \cite{Goldkuhl98}. 
Methods typically provide guidance for problem solving or for performing complex tasks and have a prescriptive character. 
Methods comprise several activities\cite{Goldkuhl98} that can be performed in parallel and allows for optional activities in some situations. 
A comprehensive method description contains the following elements:
perspective, framework, cooperation principles and all \textit{method components}.  The \textit{framework} element describes the individual method components and their relationships, sequencing and conditions under which they need to be carried out. 
The \textit{cooperation principles} are described in terms of responsible roles to carry out tasks and their skills, the way the roles cooperate and responsibility assignments for the tasks or for method components are made.
The \textit{perspective} from which a method is described influences what is considered important and is often related to the aims and purpose of the method.
The method components are concepts (artifacts and results relevant for the method), procedure (the process of identifying the artifacts) and notation (for capturing the artifacts).
The first step when creating a new method is to  analyze the state-of-the-art using the conceptualization of \cite{Goldkuhl98}, which we will present next. We will describe the new method, called AdProv, in \autoref{subsec:method}.

\subsection{ State of the Art Analysis}
We summarize the status of all method elements in \autoref{table:analysis} based on the state of the art presented earlier (Section \ref{sec:literature}).

\renewcommand{\arraystretch}{1.2}
\begin{table}
\centering
\caption{Method State of the Art Analysis - Summary}\label{table:analysis}
\makebox[1 \textwidth][c]{%
\resizebox{1.0 \textwidth}{!}{%
\setlength{\tabcolsep}{0.11cm}
\begin{tabular}{lp{0.65\textwidth}}
\toprule
\textbf{Method element} & \textbf{Status} \\
\midrule
Perspective%
& Only parts of what the method has to do are known \\

Framework%
& So far only some disconnected parts of the process are defined: capturing and visualization of workflow provenance in scientific workflows, logging event logs in BPM \\

Method components%
& Not explicitly and completely defined. Some components defined for scientific workflows targeting provenance of workflow execution. Some components defined for capturing adaptations in business workflows \\

Procedure%
& Partially defined procedures, not as part of a method definition (section \ref{sec:literature})\\

Concepts%
& Some concepts defined in several scientific publications (section \ref{sec:literature}) \\

Notation%
& Some notations known and available –BPMN, XES, PROV-O Extensions to XES for change events, data models for capturing adaptations not available yet.\\

Cooperation Principle%
& Cooperation roles partially defined in literature but not as part of any method. \\

\bottomrule
\end{tabular}
}
}
\end{table}

We can clearly identify the lack of a comprehensive method for capturing, collecting, storing, retrieving and visualizing the provenance of workflow adaptations. The conclusion from this method analysis step is that there is no available description of the method perspective, neither is there a complete framework definition. 
We also note that some of the method components are available from either the scientific workflows or BPM fields, but no connection has been established between them nor has a common motivation or purpose been identified. 
Some of the available notations that could be used for expressing some of the concepts need to be extended to render adaptations first class citizens.
Cooperations and corresponding roles have been assigned to software components/agents/tools in literature for specific scenarios only.

\subsection{Method Description}
\label{subsec:method}

The method description will follow the standard elements defined above. 

\textbf{Perspective}: The goal of our method is to ensure the provenance of adaptations of (running) processes/workflows. Therefore the method aims at defining the steps that need to be carried out to capture provenance of adaptations, store the provenance information and allow for retrieval and visualizations of that information.
The method will allow business and scientific workflows experts to document changes made and help them understand these changes and potentially why, by whom, and what the impact of these changes was on the running instances.
Process systems developers will use the method to design and implement and/or integrate the necessary software components/agents/services for capturing provenance of adaptation in different scenarios.

\textbf{Cooperation}: As we aim at automating the method realization, the cooperating participants in the method are software components, also called software agents. These are the general purpose information systems or WfMS producing/emitting the process execution events, the software used to detect changes in the process events and produce the change events, the \ph{}, the information systems retrieving the provenance information of processes or of their adaptations only, the visualization tools.

\textbf{Method Components}:
The components of the AdProv method are presented in Table \ref{table:method-components}. Consistent with the state-of-the-art overview, some of the concepts, procedures and/or notations are already available from the research in scientific workflows and BPM. The concepts that are still missing are clearly identified in the table. We will introduce de design of the missing concepts in section \ref{sec:architecture}.

\begin{table}
\centering
\caption{AdProv Method Components}\label{table:method-components}
\makebox[1 \textwidth][c]{%
\resizebox{1.0 \textwidth}{!}{%
\setlength{\tabcolsep}{0.11cm}
\begin{tabular}{lp{.4\textwidth}p{.45\textwidth}}
\toprule
\textbf{Concept} & \textbf{Procedure} & \textbf{Notation} \\
\midrule

Process model (diagram)%
& Created by business analyst or scientist. Models all workflow dimensions%
& BPMN (or other notation)\\

Process (execution) events%
& Created by the IS carrying out the processes%
& Any representation used by the IS\\

Change events (\textbf{defined here})%
& Created by either IS executing the processes or derived (as defined in the method)%
& Extended XES (or other formats) \\

\makecell{Process logs (needed for process \\mining/anomaly detection etc.)}%
& Created by the IS carrying out the processes%
& XES (or other formats) \\

\makecell{Process instance state/process trace\\(for the comparison case)}%
& Created by the IS carrying out the processes%
& XES (or other formats) \\

\makecell{Data model for\\ complete provenance information}%
& Designed by Process System developers%
& Any notation for data modelling (UML, ER Diagrams, PROV-O, etc.) \\

\makecell{Data model for\\ Adaptation provenance information\\ (\textbf{defined here})}%
& Designed by Process System developers%
& Any notation for data modelling (UML, ER Diagrams, PROV-O etc.) \\

\makecell{Provenance Holder service architecture\\ (\textbf{defined here})}%
& Designed by Process System developers%
& Any standard notation for software architecture and integration \\

\makecell{Visualization tool and mapping model \\(\textbf{defined here})}%
& Designed either by Process System developers or Integrated%
& Existing Visualization tool for PROV-O or any other tool as desired to use by the users.\\

\bottomrule
\end{tabular}
}
}
\end{table}

\begin{figure}[!h]%
    \centering
    \includegraphics[width=1\textwidth]{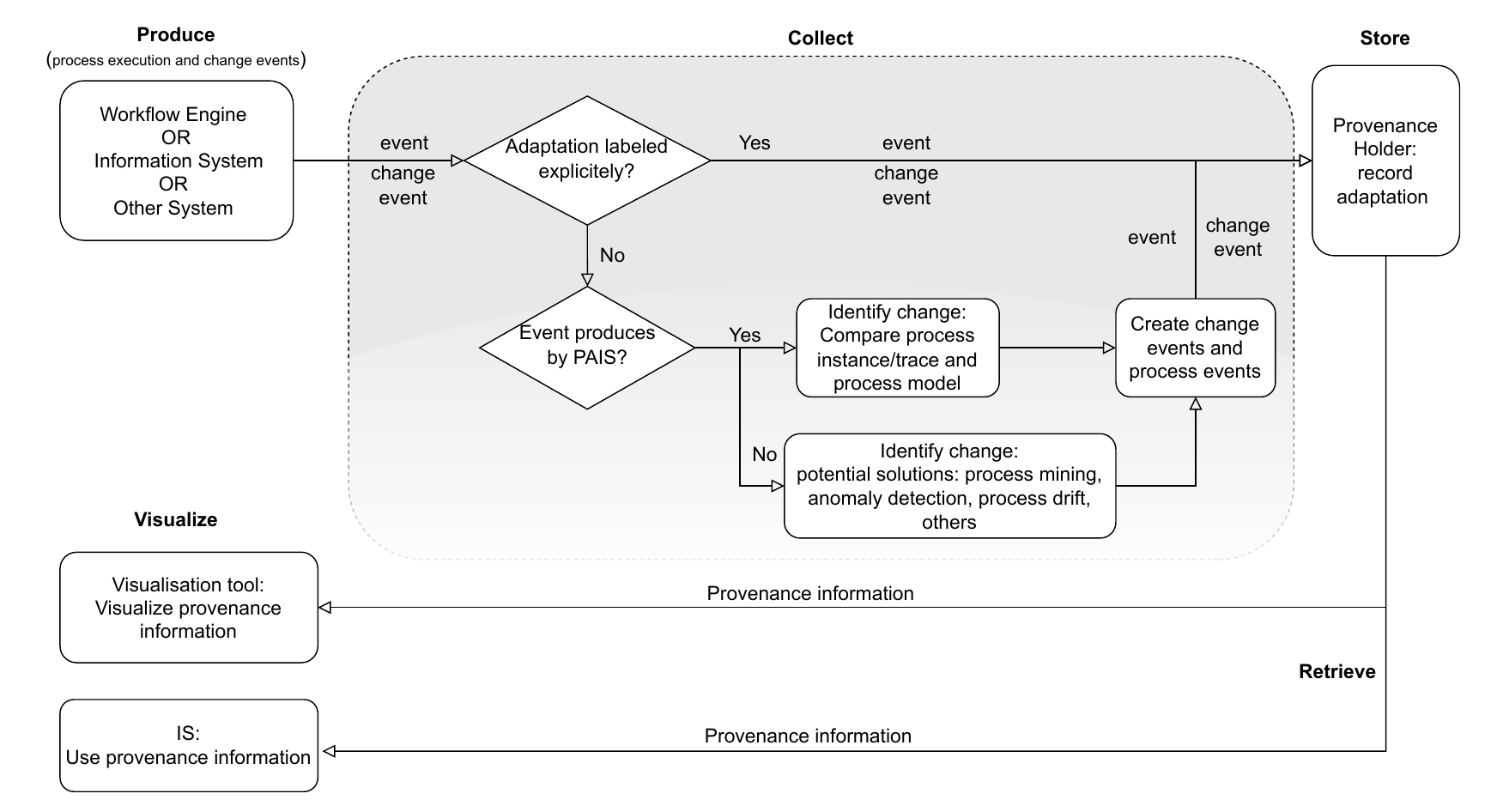}
    \caption{AdProv method steps}
    \label{fig:method}
\end{figure}

\textbf{Framework}:
The AdProv method has five high level phases/steps: produce process execution events, collect provenance information, store provenance information, retrieve and visualize. 
The relationships between the components and the steps of the method are depicted in Figure \ref{fig:method}.

The first step of the method \textbf{Produce}s the process events that reflect the execution state of processes. The process events can be produced by a workflow engine, a general-purpose IS (Information system) or any other (software) system that captures the \textit{process execution events} and can provide them as output. 
Furthermore, the adaptations in process instances should be present in the process execution events explicitly or implicitly for the method to work. We distinguish between process execution events and process change events (see \autoref{fig:method}). \textit{Change events} contain more information than the standard process execution events and capture information about the change type, the initiator of the change and other (optional) information like the impact of the change, when it was created and so on. More details on the change events are presented in \autoref{sec:datastructuresandmodels}.

The goal of the \textbf{Collect} step is to prepare the execution and change events for storage by the \ph{} service. Depending on the source of the produced process events and if they are explicitly representing a process instance change or not there are two alternative ways to carry out the collection. If the \textit{adaptations are explicitly labeled} in the process event, then the \ph{} can consume this input directly and store the information about that change. A translation step (not presented in the figure) might be necessary to convert the process event format into the internal data model used by the \ph{}.
If the \textit{process events do not contain any information about the changes} that have been carried out on the process instance then the changes need to be identified first. We consider two options here:  1) If the process events are produced by a PAIS and the process execution is governed by a process model, then the way to identify changes is to compare the process instance state/trace with the process model and detect the changes. 
Such approaches are available from BPM research and any of them can be applied (e.g. \cite{waas}). 2) If the producer of the process events is not a PAIS or only process logs are available, then the changes can be identified using approaches known from process  mining, anomaly detection and process drift \cite{Stefanie/ChangeMininginaPMS,usingprocessmining,Alirezaei2020}. 

After the changes have been identified, they need to be translated into the format used as input to the \ph{}. 

During the \textbf{Store} step all process execution events and process change events are stored as provenance information. The data model used for that is a matter of design decision. We recommend using a data model closer to the PROV-O standard for provenance representation.

The \textbf{Retrieve} step of AdProv is triggered by a user, who wants to use the provenance information. The retrieval can target one process instance or many. 
The data model of the return result could be close to the one used for storage or another one; the latter case will require one more translation sub-steps.

The last step is the \textbf{Visualization} of the provenance information of adaptations or the whole process execution, including the adaptations. This is based on the provenance information retrieved. The consumers of this information are either visualization tools or IS which may use it for further processing.

\section{Method Realization}
\label{sec:realization}
In this section we introduce the missing concepts and related procedures of the AdProv method that were identified in the section \ref{sec:method}. The concepts and techniques available from previous work are used when needed.
\subsection{Data Model for Change Events}
\label{sec:datastructuresandmodels}
Our method assumes that the process execution events are produced by a process execution environment.
We assume that such events are in XES\footnote{\url{https://www.tf-pm.org/resources/xes-standard}}, which is the current standard format and  it allows for custom extensions.

\begin{figure}[!h]
    \centering
    \begin{subfigure}[t]{\textwidth}
        \centering
        \includegraphics[width=.7\textwidth]{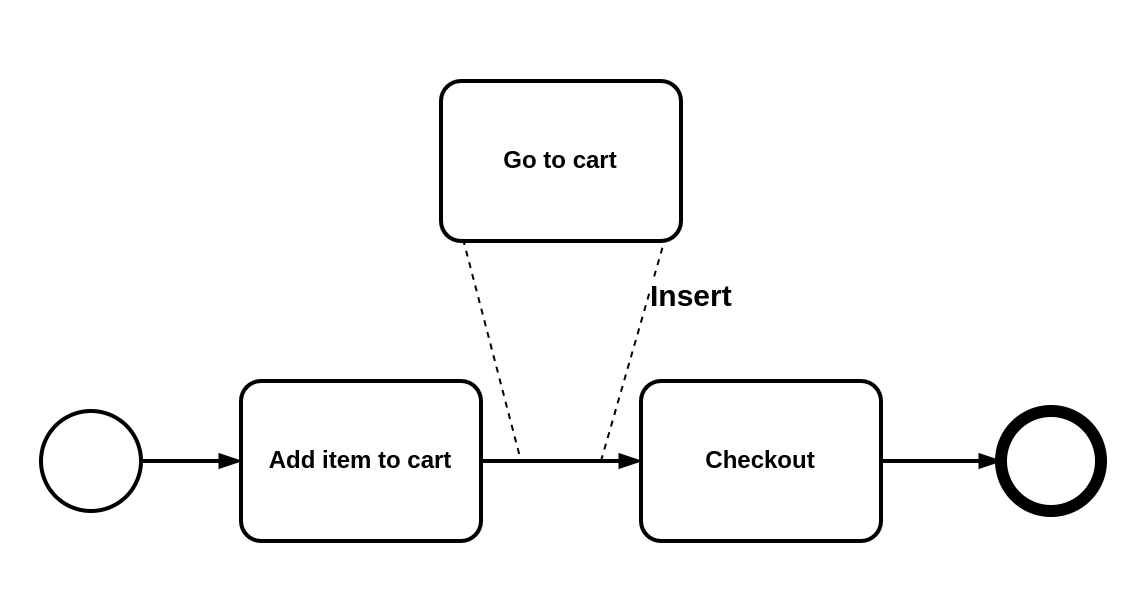}
        \caption{Simple online shopping workflow with activity insertion}   
    \end{subfigure}
    \hfill
    \begin{subfigure}[b]{0.5\textwidth}
        \centering
        \includegraphics[width=0.9\textwidth]{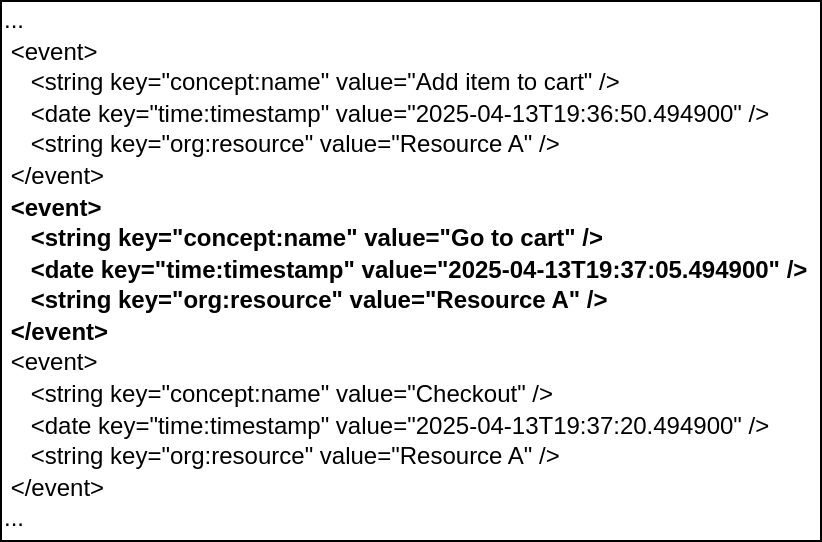}
        \caption{XES log with inserted activity}
    \end{subfigure}%
    \hfill
    \begin{subfigure}[b]{0.5\textwidth}
        \centering
        \includegraphics[width=\textwidth]{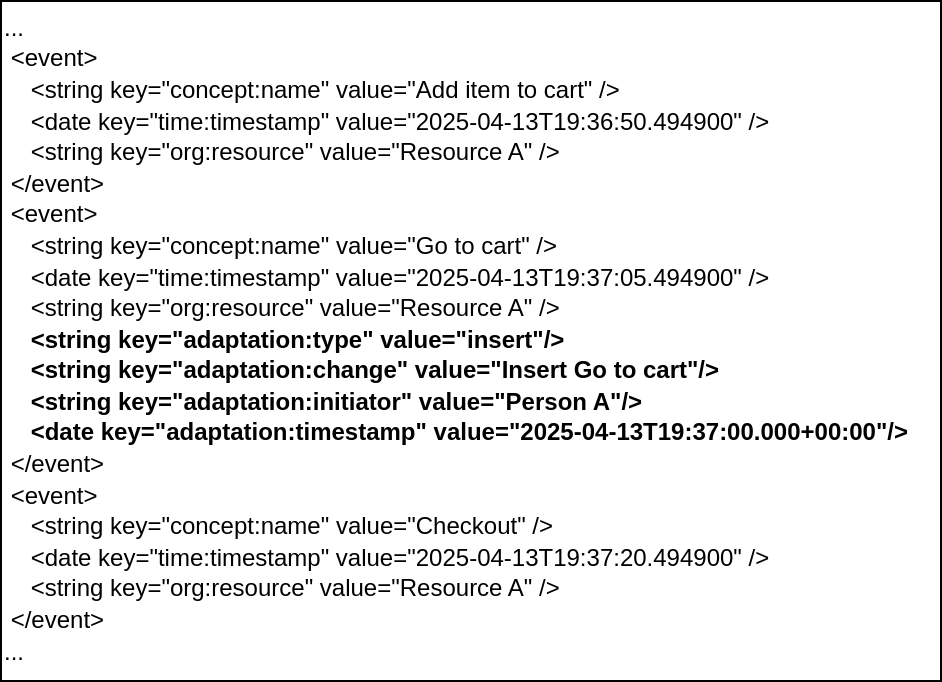}
        \caption{XES log with adaptation extension}
    \end{subfigure}
    \caption{Simple online shopping workflow with XES log examples}
    \label{fig:log-events-to-provenance-info-xesext-int}
\end{figure}

In order to be able to detect and subsequently record adaptations in running workflow instances it is necessary to enrich the given event information and to provide the supporting data structures for \textit{change events} (see Figure \ref{fig:method}).
These change events need to contain the type of adaptation (insert, delete, etc.), the location (where), the adaptation itself (what), the initiator (who) and the timestamp (when) of when the change has been made (not when it has been executed  - this is information related to the execution of the change).For that purpose we defined an XES extension - the  Adaptation XES extension\footnote{\url{https://raw.githubusercontent.com/ProvenanceHolder/ProvenanceHolder/refs/heads/main/adaptation.xesext}}
(cf. \autoref{fig:log-events-to-provenance-info-xesext-int}) that allows for capturing the additional adaptation related information.

AdProv requires therefore the process execution environment to produce the change events and in our realization of the method we assume that both execution and change related information is available in the event streams produced.

We only record two types of control flow adaptations -- insert and delete -- as it is postulated by process adaptation patterns \cite{WeberReichertonAdaptationBook} that all process changes/adaptations can be broken down into  combinations of insert and delete.

In \autoref{fig:log-events-to-provenance-info-xesext-int} (c) we show an example of the use of the change events and the \textit{Adaptation XES extension}. The process log of an adapted process instance will contain the execution events of  the original process models, the execution events of the inserted activity (\enquote{Go to cart}) and the additional information about the type of change, initiator and time of making the change. 

If the process execution environment does not produce such an event stream, AdProv states that the change events need to be created using an alternative change identification step in the Collect phase of AdProv (see \autoref{fig:method}).

The events are then recorded and stored by the \ph{} in a data structure resembling the PROV-DM \cite{provdm-w3c}.

\subsection{\ph{} Architecture}
\label{sec:architecture}
We identified the \ph{} service as one of the missing components of AdProv and in this section we will give an overview of its architecture. 

\begin{figure}[htp]
    \centering
    \includegraphics[width=.95\textwidth]{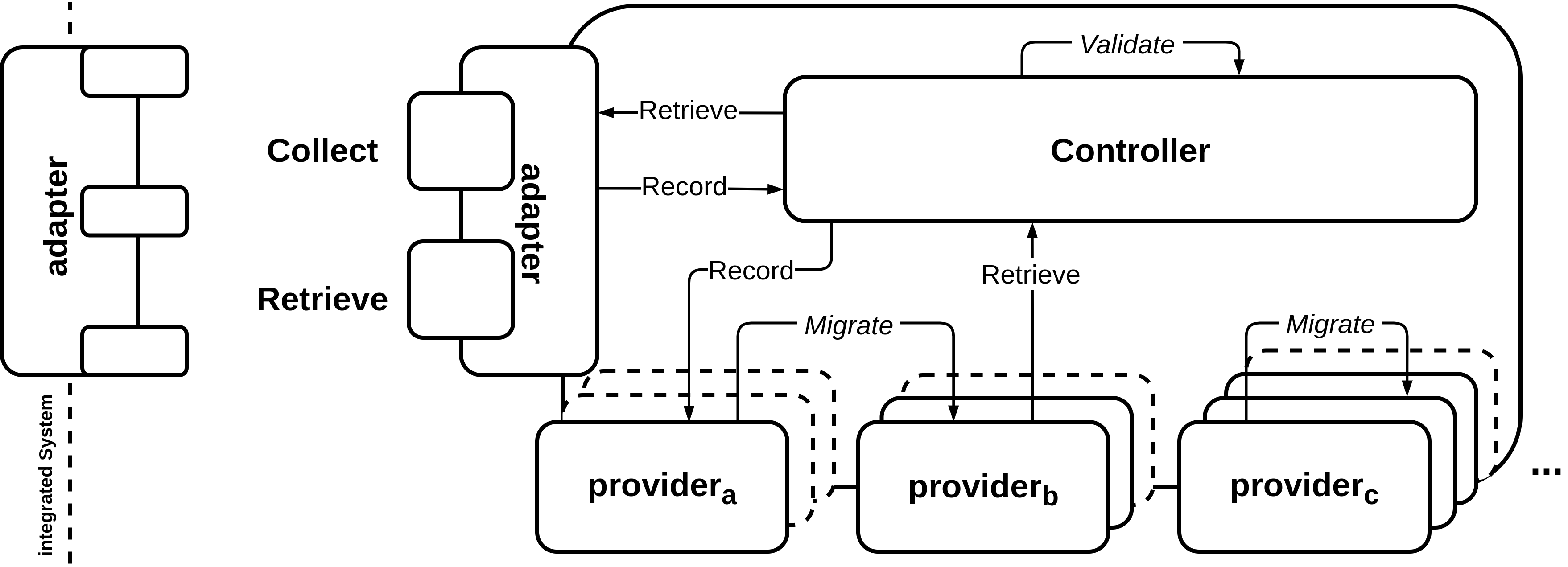}
    \caption{\ph{} Architecture components; adapter, external operations and internal methods (adopted from \cite{EDOC23})}
    \label{fig:arch}
\end{figure}

The \ph{} components are: the controller, the adapter and one or more provenance providers (\autoref{fig:arch}).
The adapter is the interface of the service providing \textit{two main external operations}:
1) \textit{\textbf{Collect} provenance data} 
and 2) \textit{\textbf{Retrieve} provenance information}.%
The components carry out four \textit{interaction scenarios} in order to realize the two external operations of the \ph{} service.
The interaction scenarios are always combinations of several of the internal methods\footnote{The term \textit{method} is used for disambiguation purposes only.};
the (internal) methods are: \textit{Record}, \textit{Retrieve}, \textit{Validate} and \textit{Migrate}.
The \textit{adapter} is the component ensuring the \textit{integration} of the \ph{} with other systems. In order to do so, the \textit{adapter} is split into two parts:
a) the \ph{}'s interface, providing the two external operations: \textit{Collect} and \textit{Retrieve} and b) the integrative part residing on the side of the system it is to be integrated with.
By doing so we have both a standardized interface, i.e. service, and customizable part to be able to integrate with existing systems and environments.

\textit{Provenance providers} (short: \textit{providers})
have to implement three 
methods, i.e. \textit{record, retrieve and migrate} and ultimately store the provenance information.
The employed storage technology as well as the possible needs of different types of workflows have an impact on implementation characteristics and complexity.

The \textit{controller} is in charge of the interaction between all components as combinations of the four methods into the realization of the two external operations. 
The \textit{Collect operation} is a combination of the receive, validate and record methods thus storing the the  provenance information on a provider.
The \textit{Retrieve operation}, combines the methods retrieve and validate. For more details on the architecture\footnote{\url{https://github.com/ProvenanceHolder/ProvenanceHolder}}
please consider \cite{EDOC23}.

\subsection{Adaptation Visualization}
\begin{figure}[htb!]
    \centering
    \includegraphics[width=0.99\textwidth]{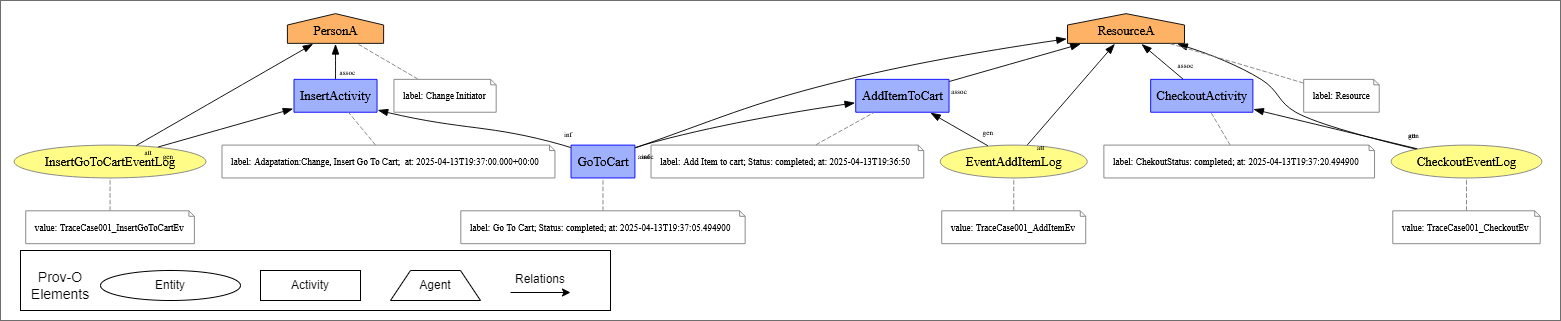} 
\caption{Provenance graphs generated based on the PROV-O specification for the example in Figure\ref{fig:log-events-to-provenance-info-xesext-int}}
    \label{fig:exampleprv}
\end{figure}

In order to ensure semantic interoperability the provenance data will be retrieved in a representation following the standard provenance ontology PROV-O and data model PROV-DM \cite{provdm-w3c}. The Visualization step of AdProv can be implemented using any tool that is compliant with this ontology. The provenance representation of the running example is shown in Figure \ref{fig:exampleprv}. We implemented a visualization with ProvToolbox\footnote{https://lucmoreau.github.io/ProvToolbox/}.

For the visualization we mapped elements of the event logs to elements of the Prov-O ontology: process activities and changes are mapped to Prov-O Activities, Resources in process instances are represented by the Agent elements in Prov-O, and Prov-O Activities are actions that write into the event log elements, called Entities in Prov-O. 
The example provenance visualization hence shows that three process instance activities have been carried out: \textit{AddItemTo Cart}, \textit{GoToCart} and \textit{Checkout}, as well as the fact that the \textit{GoToCart} process activity has been inserted at runtime. This fact is visualized using a relation between the \textit{GoToCart} and \textit{InsertActivity} elements of Prov-O. 
All three process activities are carried out by the resources \textit{ResourceA}, as stated in the event log, whereas the insertion is carried out by the resource \textit{PersonA}. Labels are associated to each element to depict relevant attribute values and timestamps, that are also present in the event logs.

\section{Feasibility Discussion}
\label{sec:feasibility}

The feasibility of the AdProv method can be illustrated on basis  of the example workflow shown in
\autoref{fig:log-events-to-provenance-info-xesext-int} (a) depicting a simple online shopping workflow with two activities (\enquote{Add item to cart} and \enquote{Checkout}). If during the execution of one instance of this workflow model an adaptation is needed and, for example, the activity \enquote{Go to cart} is inserted, this ad-hoc adaptation has to be captured in the provenance information of that workflow instance.
A WfMS using the AdProv method will have to be able to generate the execution events for this workflow instance in XES format following the Adaptation XES extension we introduced here. This way the change event will be captured in the provenance information as shown in \autoref{fig:log-events-to-provenance-info-xesext-int} (b). With that the \textit{Produce} and \textit{Collect}  steps of AdProv are completed. As the Adaptation XES extension is a standard-based one it is realistic to assume that adaptive execution environments that can produce XES logs can be extended to generate the change events as well. Subsequently, the change event reflecting the insertion of the activity \enquote{Go to cart} will be \textit{Stored} by the \ph{}. Our current implementation of the \ph{} service allows for storing execution and change events in a data mdoel similar to the Prov-O representation of provenance data. However, other representations can be used in alternative implementations of the \ph{} as long as a mapping to the XES event logs is made available. 
When a user wants to inspect the provenance information for the same process instance, the \textit{Retrieve} step of AdProv takes place. For that the "Retrieve" operation of the \ph{} has to be invoked by either a Visualization tool or another information system - this is a design decision and any choice will be suitable as long as the Prov-O presentation of the provenance information is a valid input. One such freely available tool is ProvToolbox, which we used with the running example (Figure \ref{fig:exampleprv}). It remains to be investigated what the most suitable visualization is for different domains. As a result we observe that using the  AdProv method, collecting and retrieving the provenance of workflow adaptations is achievable with reasonable implementation and integration effort.

In case that the information system executing the workflows cannot produce event logs with change events, then the Collect step of AdProv will follow the alternative path through which the event stream will be extended with the change events. As there are many realizations of approaches from process mining, in particular anomaly detection and change mining, that can detect runtime adaptations, the AdProv method allows for their reuse. The rest of the steps of the method will be carried out as described above. This scenario requires more implementation and integration effort than the previous one, however it might be the only option in case the process execution environment is not possible to extend with the functionality to generate the change events.

\section{Concluding Remarks}
\label{sec:conclusions}
In this paper we address a gap in the state of the art in workflow management research  that is the lack of a method for capturing provenance information of runtime workflow adaptations/changes. 
We fill the gap by introducing the AdProv method that comprises five steps: produce executions and change events, collect, store and retrieve change and execution  provenance information, and visualize the provenance of workflows and their adaptations.
If the process execution environment can create the provenance information of adaptations in the form of change events in the process event streams, an XES extension we defined in this work can be used as a standard way to exchange the information.  For the storage of the information we use a straightforward procedure of mapping the change events into a provenance model. We use the PROV-O standard for this purpose and defined a mapping between the event stream and PROV-O. 
For the case that the process execution environment cannot generate the change events, existing approaches from process mining, anomaly detection and process drift detection for detecting changes in process event streams (i.e. changes are not explicitly recorded as such or not labeled) can be reused, so that these changes can also be captured as provenance information.
We also provide an architecture  and a first implementation of the \ph{} service that realizes the Collect and Retrieve steps of the method. We consider the AdProv method for building systems capable of collecting and retrieving provenance of workflow adaptations feasible, as its goals are achievable with reasonable implementation and integration effort. 

The method provides an overall view of the interaction between the process execution environment, the \ph{} service and the visualization tools and clearly identifies the interfaces that will facilitate the integration of all components. It also shows the alterative approaches for realization of a modular but complete system that can record provenance information of adaptive workflows. The method highlights the potential reuse of existing approaches from process mining, in particular anomaly detection and change mining. Using the method, system architects can derive the requirements on their existing systems and extensions they need to carry out to be able to implement complete coverage of workflow provenance that accounts for adaptations.
In future we plan to refine the method to allow for provenance of collaborative adaptations spanning several processes and we will focus on trusted collaborative adaptations, that would prove to participants in a collaboration that changes have been carried out without disclosing the complete execution information.
\begin{credits}
\subsubsection{\ackname} 
This study was partly funded by the EU Chips Joint Undertaking project AIMS5.0 and the Dutch National Funding Agency RVO under grant agreement number 101112089. 
\end{credits}
\bibliographystyle{splncs04}
\bibliography{ref}
\end{document}